\documentstyle[psfig]{mn}
\title{Is the Galactic Disk older than the Halo?}
\author[Giovanni Carraro et al.]
       {
Giovanni Carraro$^{1}$, L\'eo  Girardi$^{2}$, and Cesare Chiosi$^{1,3}$\\
       $^{1}$ Dipartimento di Astronomia, Universit\`a di Padova, Vicolo
dell'Osservatorio 5, I-35122 Padova, Italy \\
       $^{2}$ Max Planck Institut f\"ur Astrophysik,
       Karl-Schwarzschild-strasse 1, D-85748, Garching bei M\"unchen,
Germany\\
       $^{3}$ Visiting Scientist, Max Planck Institut f\"ur Astrophysik,
       Karl-Schwarzschild-strasse 1, D-85748, Garching bei M\"unchen,
Germany\\
E-mail: {\tt carraro,chiosi@pd.astro.it; leo@mpa-garching.mpg.de}
}

\date{\it Submitted: November 1998}
\pubyear{1998}
\begin{document}
\maketitle
\title{The age of the Galactic Disk}

\begin{abstract}

Aim of this study is to infer the age  of the Galactic Disk by means
of  the ages of old open clusters, and comment on some recent
claims that the Galactic Disk can be older than the Halo.
To this purpose, we analyze the Color--Magnitude Diagrams (CMDs) of six
very old open clusters, namely NGC~188, NGC~6791, Collinder~261,
Melotte~66, Berkeley39 and Berkeley~17,  and   determine  
their ages. 
For each cluster we use the  most recent  photometric and
spectroscopic data and metallicity estimates.
The ages are derived from  the isochrone fitting method using 
the stellar models of the Padua library (Bertelli et al. 1994,
Girardi et al 1998).
We find that the ages of these clusters fall in the range 
4 to 9-10 Gyr: Melotte~66 is the youngest whereas  NGC~6791 and Berkeley~17 
have ages of about 9-10 Gyr. Previous estimates for Berkeley~17
indicated an age as old as $12^{+1}_{-2}$ Gyr, almost falling within the
range of classical
globular clusters.  In our analysis, this cluster
is always very old but perhaps somewhat younger than in previous
studies. 
However, we call attention on the fact that the above ages are
to be taken as provisional estimates, because of the many uncertainties
still affecting stellar models in the mass range 1.0 to 1.5
$M_{\odot}$.
Despite this drawback of extant theory of stellar structure,
if NGC~6791 and Berkeley~17 set the  limit to the age
of the Galactic Disk, this component of the Milky Way  can be as old as
about 9-10 Gyr, but surely younger than the Galactic Halo, at least  as
inferred from   recent  determinations of the age of  globular
clusters.
Finally, it is worth recalling that 
 open clusters can only provide a lower limit to the age of
the Galactic Disk, while other indicators - like White Dwarfs - 
are perhaps more suited to this task.
\par
\end{abstract}


\section{Introduction}
In the last decade considerable efforts have been made to collect
good quality photometric and spectroscopic data for the oldest open
clusters in the Galactic Disk (see Friel 1995 for an exhaustive   
review on the subject), with the aim of establishing constraints on 
the age of this component of the Milky Way and to assess whether the
age distribution of the open clusters overlaps that of globular
clusters (cf. Janes \& Phelps 1994).

Since Cannon (1970), the taxonomy of the population of old open
clusters  is generally based  on the morphology of their CMD.  
Intermediate age open clusters
show the so-called Hertzsprung gap between the main sequence band
and the red giant region,  and a populous clump of  He--burning
red giant stars, the analog of the Horizontal Branch in globular clusters.
Prototypes of this class are NGC~752, NGC~7789 and NGC~2158.
At increasing age, the separation  between the main sequence and the
red giant region
gets smaller, and  the red clump gets scarcely 
populated. M~67 and NGC~188 are
two  well studied examples of this class.

For the purposes of this study, we extend the above classification
to   three groups: (i) the intermediate-age clusters,
i.e. those whose age falls in the interval defined by the
Hyades (625 Myr, cf. Perryman et al. 1998) 
and IC~4651 (1.7 Gyr, cf. Bertelli et al. 1992); (ii) the  old clusters
whose age is younger than that of M~67
(4.0 Gyr, cf. Carraro et al. 1996); (iii) the
 very old  clusters, i.e  older than M~67 and younger than the bulk of
 globular clusters (12-13 Gyr, Gratton et al. 1997).

According to Phelps \& Janes (1994), there are about 20 clusters older
than M~67, even though the published photometry is not yet 
sufficiently accurate to prove
that all of them really belong to the third group in our classification.

\begin{table*}
\tabcolsep 0.35truecm 
\caption{
Basic properties for the open clusters in our ample.\protect\\
$^{a}$Source of spectroscopic data: (1) Friel et al. (1995); 
(2) Friel \& Janes (1993).\protect\\
$^{b}$Source of photometric data: (1) McClure \& Twarog (1977); (2) Von
Hippel \& Sarajedini (1998);
 (3) Kaluzny \& Rucinski (1995) ;
 (4) Mazur et al. (1995); (5) Gozzoli et al. (1996); 
(6) Anthony-Twarog et al. (1979); (7)
Kassis et al. (1997); (8) Kaluzny (1994);(9) Phelps (1997).
} 
\begin{tabular}{ccccccccc} \hline
\multicolumn{1}{c}{Cluster} &
\multicolumn{1}{c}{$\alpha(2000.0)$} &
\multicolumn{1}{c}{$\delta(2000.0)$} &
\multicolumn{1}{c}{$l$} &
\multicolumn{1}{c}{$b$} &
\multicolumn{1}{c}{$[Fe/H]$} &
\multicolumn{1}{c}{$\sigma_{[Fe/H]}$} &
\multicolumn{1}{c}{Source$^{a}$} &
\multicolumn{1}{c}{Source$^{b}$} \\
\hline
NGC~188       &00:39.4 &  +08:04  &122.78  &  +22.46 &$-$0.05 &0.11  &(1) &(1,2)\\
NGC~6791      &19:19.0 &  +37:45  &70.01   &  +10.96 &  +0.19 &0.19  &(2) &(3)\\
Collinder~261 &12:34.9 &$-$68:12  &301.69  &$-$05.64 &$-$0.14 &0.14  &(1) &(4,5)\\
Melotte~66    &07:24.9 &$-$47:38  &259.61  &$-$14.29 &$-$0.51 &0.11  &(2) &(6,7)\\
Berkeley~39   &07:44.2 &$-$04:29  &223.47  &  +10.09 &$-$0.31 &0.08  &(2) &(7)\\   
Berkeley~17   &05:17.4 &  +30:33  &175.65  &$-$03.65 &$-$0.29 &0.13  &(1) &(8,9)\\   
\hline
\hline
\end{tabular}
\end{table*}

Despite their small number, these clusters are very interesting for
several reasons.  Firstly, they
 can be used to study the formation and early evolution -- both
chemical and dynamical -- of the 
Galactic Disk (Carraro et al. 1998a). Secondly, they can map  the
outer structure of the Galactic Disk, as they are preferentially
located high on the galactic plane and towards the galactic anticenter.
Finally, once    these clusters are accurately 
ranked in  age, they can be used to
set  a lower
limit to the age of the Galactic Disk (Sandage
1989, Carraro \& Chiosi 1994), to constrain the evolution of the Milky
Way, and to highlight 
 the parental relationship between  Galactic Disk and  Halo.

  NGC~6791 has long been considered as the oldest open cluster, to 
which Janes (1989) assigned an age of about 12 Gyr.
However more recent determinations of the age by 
Carraro \& Chiosi (1994), Carraro et al.  (1994a), and Garnavich et al. (1994)
have decreased this value down to about 
about 9 Gyr.

In the meantime another very old cluster has been discovered, namely 
Berkeley~17, whose age is estimated $12^{+1}_{-2}$ Gyr (Phelps 1997).

As far as  Lynga~7 is concerned, that was long considered as another
 very old candidate (Janes \& Phelps 1994),
it has been proven by Tavarez \& Friel
(1995) to belong to the family of Thick Disk globular clusters.

Since the bulk of globular clusters seem to be 12-13 Gyr old (Gratton et
al. 1997), the possibility  
arises that the Disk might be as old as the Halo, or even older than
this  (see the
discussion in Phelps 1997).
This fact bears very much on the time scale
of the halo collapse, the initial stages of the formation of the
Galactic Disk, and the star formation history of the Galaxy.

To cast light on this topic of paramount importance, 
we have selected 6 very old open clusters (NGC~188, NGC~6791,
Collinder~261, Melotte~66,
Berkeley~39 and Berkeley~17) for which good photometric
and spectroscopic data were available, and have derived their age in a
homogeneous fashion.
 
With respect to a similar study by  Janes \& Phelps (1994),
the situation is now much improved firstly because  better data 
are at our disposal, and secondly 
the metal content
of these clusters is known from spectroscopic determinations.
Therefore another attempt to rank these clusters as a function of the
age can be undertaken.

There is one point to be kept in mind from the very beginning, i.e.
that old and very old open  clusters have  their turn-off mass 
in the risky mass interval, in which a number of important physical 
facts occur in the interiors of their stars.  The turn-off mass 
falls in fact in the range 0.9 to about 2 $M_{\odot}$.  In this mass
range  firstly 
core H-burning switches from radiative to convective regime
(the border-line  is at about 1.0-1.1 $M_{\odot}$). Secondly, 
the transition occurs from pp-chain to CNO-cycle.
The typical mass at which this transition occurs is about 1.3-1.4 
$M_{\odot}$. 
Even more important, in stars of this mass,
the time scale at which CNO-cycle goes to equilibrium is sizable as 
compared to the core H-burning lifetime itself. As a consequence of
it, the mass size of the convective core  starts
small, grows to a maximum, and then shrinks  as a function of the age. 
At higher masses this does no longer occur because the 
time scale for CNO-equilibrium is short as compared to core H-burning 
lifetime: the convective core almost immediately reaches the maximum 
size and then gradually shrinks in mass at proceeding evolution.
Finally, at even higher initial masses, say about 2.2 
$M_{\odot}$ with solar composition,
the ignition of central He-burning switches from degenerate
to non degenerate conditions.  This mass is otherwise known as 
$M_{\rm Hef}$. All of the above limit masses, $M_{\rm Hef}$
in particular, depend on the
initial chemical  composition and treatment of convection. 
For instance the 
$M_{\rm Hef}=2.4 M_{\odot}$ solar case value,  can be as low as 1.6-1.8 
$M_{\odot}$ in presence of convective overshooting  (see Chiosi et al. 
1992 for a recent review of the subject). 

In addition to that, in those 
stars in which the convective core first increases and then decreases 
during the core H-burning phase, modeling this initial expansion of
the
convective core  is highly
uncertain even in standard schemes of convection, and the situation 
is even more difficult to deal with 
if one allows for
the occurrence of convective overshoot. 

In this latter case,
key questions to be answered 
are: firstly which kind of convective overshooting is best suited
to real stars, secondly whether overshooting immediately grows to full 
amplitude or it
regulates itself as the convective core grows, being overshooting
small when the core is small, and settled to full amplitude only when
the
core has grown to its maximum extension. A complete report on these
matters is beyond the scope of the present paper. 

The above
considerations have been made to remind the reader that assigning the
age is to some extent easier for young open clusters and very old
globulars
where the physical structure of the turn-off stars is much more clear. 
The intrinsic uncertainty and ambiguity in age assignment to old open
clusters will stand out  clearly when  discussing the CMD of
NGC~6791 and Berkeley~17.
It will be shown indeed that determining the age of old open clusters is
still a cumbersome affair awaiting for and demanding better understanding of 
stellar structure in the mass range 1-2 $M_{\odot}$.

The plan of the paper  is as follows.
In Section~2 we describe the stellar models in usage here,
the comparison between two
sources of isochrones, i.e.  Bertelli et al. (1994) and
Girardi et al. (1998), and finally the method we have adopted
to derive cluster ages, basically isochrone fit. 
Section~3,  devoted to the analysis of each 
cluster of our sample, presents the CMDs and our isochrone fits.
Finally, 
Section~4 deals with the age of  the Galactic Disk, and draws some
concluding remarks.

\section {Stellar models and isochrones }

The Bertelli et al. (1994) library of  isochrones and companion
stellar models has long been used in a large variety of astrophysical
problems going from studies of CMD of single clusters or complex
stellar
populations to spectro-photometric synthesis (Bressan et al. 1994).

Recently, Girardi et al. (1998) have revised the input physics
of the stellar models and  generated a new library of  isochrones. 

In the following we will shortly summarize the key assumptions of
the Girardi et al. (1998) models and highlight the points of major 
difference with respect to Bertelli et al. (1994).

\subsection{Stellar  models}

The stellar models are from Girardi et al.\ (1998). They
consist of a large set of evolutionary tracks for metallicities
ranging from $Z=0.001$ to $0.03$, and masses from 0.15 to 7 $M_\odot$.
Models of low-mass stars are computed at typical mass intervals of
0.1~$M_\odot$. 
Work is progress to extend  the library to other metallicities and/or
to the  range of massive stars.

{\it Details of the input physics}. All details about the input physics of 
the tracks can be found in
e.g.\ Girardi et al.\ (1996, 1998). Suffice it to mention that they
include updated OPAL (Iglesias \& Rogers 1996) and low-temperature
opacities (Alexander \& Fergusson 1994), and a revised equation of
state. In comparison with the previous set of the Padua stellar tracks
(Bertelli et al.\ 1994 and references therein), the present ones have
slightly different lifetimes (see Girardi et al.\ 1996), and giant
branches which are tipically hotter by about 50 to 100~K. These
differences result mainly from the inclusion of Coulomb interactions
in the equation of state, and from the new low-temperature opacities.

The mass and metallicity resolutions of the grids of evolutionary
tracks are suitable for the derivation of accurate isochrones in the
CMD. The method in usage here  is that of interpolating between
equivalent points along the tracks (see e.g.\ Bertelli et al.\ 1994).
In this way, we generate isochrones for any intermediate value of age
and metallicity.  Mass loss in the RGB is according to the Reimers
(1975) formula with an efficiency factor of $\eta=0.4$.  The
bolometric corrections and $T_{\rm eff}$--colour transformations are
 derived from the Kurucz (1992) library of model atmospheres
(see Bertelli et al.\ 1994 for details).

{\it Calibrating against the Sun}.  Bertelli et al. (1994) and 
Girardi et al. (1998)
calibrate the stellar models, i.e. choose the mixing length parameter
$\alpha\times H_P$, where $H_P$ is the local pressure scale height,
for the outer superadiabatic convection, following different methods.
Bertelli et al. (1994) started assuming the metallicity and helium
content  of the Sun, i.e. $Z_{\odot}=0.020$ and $Y_{\odot}=0.28$, and looked
for the value of $\alpha$ that  would fit the effective temperature
of the Sun at the canonical age of 4.5 Gyr. They got $\alpha=1.63$
but with a small offset in the Sun 
luminosity ($logL/L_{\odot}=-0.017$).
Girardi et al. (1998) prefer to start from the Sun metallicity
(Z=0.019),
luminosity and effective temperature and look for the values of
$\alpha$
and $Y_{\odot}$ that would fit the Sun at the same age. They get
$\alpha=1.68$ and $Y_{\odot}=0.273$. The small difference in $\alpha$
implies that the isochrones of the Girardi et al. (1998) library
will be somewhat hotter during the RGB phase than those of Bertelli et al. (1994).

{\it Enrichment law  $\Delta Y /\Delta Z$}. All the models calculated for the Bertelli
et al. (1994) library made use of the enrichment law 
$\Delta Y /\Delta Z=3.0$ or equivalently  $Y=Y_P+3.0\,Z$, with
$Y_P=0.23$
the primordial helium content (Torres-Peimbert et al. 1989, 
Olive \& Steigman 1995). Girardi et al. (1998) adopted
a slightly different relationship, i.e.   $Y=0.23+2.25\,Z$ with
$\Delta Y /\Delta Z=2.25$, such as to reproduce
 the initial helium content of the Sun $Y_{\odot}=0.273$ as
derived from the calibration of the solar model.

{\it Convective overshooting.}
In all the tracks with $M\ge 1.0$~$M_\odot$, some amount of
convective overshooting is adopted.  This is expressed by means of the
parameter $\Lambda \times H_P$ (see Alongi et al. 1993 for more details on the
subject). 
The  way in which convective overshooting is let develop in the range
of low mass stars (say 1.0 to 1.5 $M_{\odot}$) bears very much not only
on the properties of stellar models, but also on the associated
isochrones
and related problem of ranking ages. 

In Bertelli et al. (1994)
the following recipe was adopted: $\Lambda=0.$ for $M \leq 1.0
M_{\odot}$, $\Lambda=0.25$ for $1 M_{\odot} < M \leq 1.5 M_{\odot}$,
and $\Lambda=0.50$ for $ M > 1.5 M_{\odot}$.\\
Girardi et al. (1998) modified this recipe assuming that
$\Lambda$  increases linearly with the star mass in the interval
$1.0\le(M/M_\odot)\le 1.5$, getting to its maximum value of
$\Lambda=0.50 $ only for $M>1.5$~$M_\odot$.

\begin{figure*}
\centerline{\psfig{file=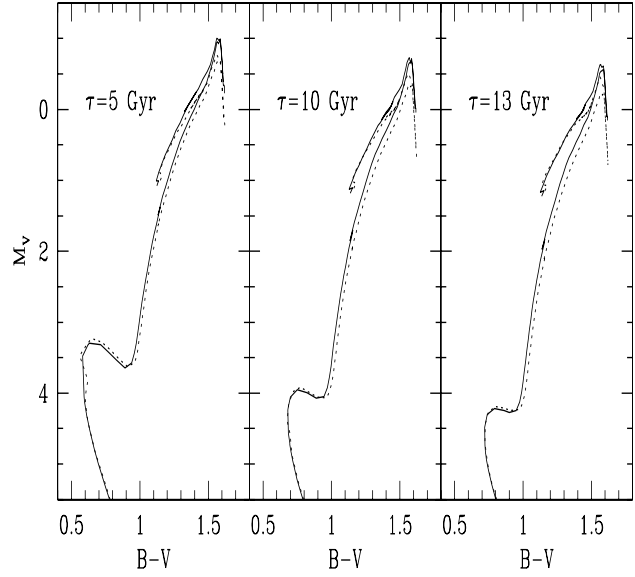,height=9cm,width=9cm}}
\caption{Isochrones of 5, 10 and 13 Gyr from Bertelli et al. (1994),
dotted lines,
and Girardi et al. (1998), solid lines. The chemical composition
is $Y$=0.28, $Z$=0.020 in Bertelli et al. (1994) and $Y$=0.273,
$Z$=0.019 in Girardi et al. (1998)}
\label{fig_isoc}
\end{figure*}

\begin{figure*}
\centerline{\psfig{file=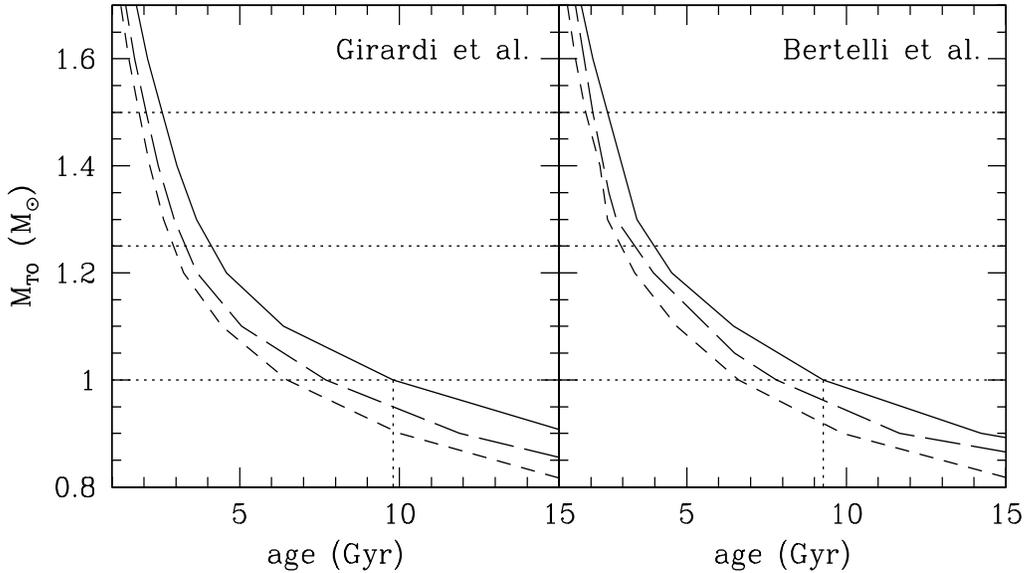,height=9cm,width=14cm}}
\caption{The turn-off mass (in solar units) as a function of the age
(in Gyr) for the metallicities Z=0.004, Z=0.008, and Z=0.02. See the
text for the meaning of the horizontal and vertical dotted lines}
\label{fig_mto}
\end{figure*}

\subsection{Comparing isochrones from different sources}

In this section we compare the  isochrones 
by Girardi et al. (1998) with those by Bertelli et al. (1994)
in order to estimate possible systematic differences in the final age
assignment caused by  using different sources of stellar models
and/or isochrones.

In Fig.~\ref{fig_isoc} we plot three groups of isochrones taken from
Bertelli et al. (1994), dotted lines, and Girardi et al. (1998), solid
lines, for the ages of 5, 10, 13 Gyr as indicated. For all cases
the chemical composition is solar, i.e. $Y$=0.280,$Z$=0.020 in
Bertelli et al. (1994) and $Y$=0.273, $Z$=0.019 in Girardi et al.
(1998). While at the
turn-off the two types of isochrones are almost indistiguishible,
their RGBs differ in color by a sizable factor. Looking at the case
of 10 Gyr, the red clump phase (stationary core He-burning) of the
Girardi et al. (1998) isochrone is fainter by about 0.15 mag with
respect to that of Bertelli et al. (1994) isochrone.

To better understand the effect of the different assumptions
for the calibration against the Sun, the enrichment law, and 
convective overshooting in particular, we plot in Fig.~\ref{fig_mto}
the turn-off mass  $M_{TO}$ as a function of the age for both sets of 
isochrones. The left panel is for Girardi et al. (1998), whereas the
right panel is for Bertelli et al. (1994). In both panels we draw 
the lines of constant mass $M_{TO}=1.0 M_{\odot}$, 
$M_{TO}=1.25 M_{\odot}$, and  $M_{TO}=1.5 M_{\odot}$.
No convective core can develop for ages older than set by the
intersection with the $M_{TO}=1.0 M_{\odot}$
line (the ages depend, however,  on the chemical composition). Girardi et al. (1998)
isochrones are less affected by overshooting with respect to the case
of Bertelli et al. (1994) for all ages 
older than set by the intersection  with the  $M_{TO}=1.25 M_{\odot}$
line. Once again they depend on the metallicity. The opposite for
all the younger
ages  corresponding to turn-off masses in the range 
$1.25 M_{\odot} < M_{TO} < 1.5 M_{\odot}$.

Particular useful for age ranking, is the so-called 
$\Delta V_{\rm TO}^{\rm RGC}$ method, which correlates the magnitude
difference between the red giant clump (RGC) and turn-off (TO) stars 
to the age. 
This relation owes its existence to the fact that the RGC luminosity
is almost age independent for ages older than about 2 Gyr. This
combined with the dependence of the turn-off luminosity on the
cluster age, makes the above difference a reasonably good age indicator.
The relation
is, however, sensitive to the chemical composition and other physical 
details of model construction. This basic calibration is presented
in Table~2 limited to the case of solar composition both for the Girardi
et al. (1998) and Bertelli et al. (1994) isochrones.
The magnitude of the RGC stars is taken
at the lowest luminosity end of the clump.

At given age, the Girardi et al. (1998)  isochrones
yield values of $\Delta V_{\rm TO}^{\rm RGC}$, that
are at least 0.15~mag smaller  than those from the 
 Bertelli et al. (1994) isochrones. In general the following
relation holds

\begin{equation}
   {\Delta (\Delta V_{\rm TO}^{\rm RGC}) \over \Delta log t } \simeq 1.9
   \quad \quad {\rm mag/dex} 
\end{equation}

The 0.15 mag difference in $\Delta V_{\rm TO}^{\rm RGC}$ between
Girardi et al. (1998) and
Bertelli et al. (1994) together with the above relationship
imply that the ages derived 
from using the Girardi et al. (1998) isochrones are about 20\% 
older  than those one would obtain from using the Bertelli et
al. (1994) isochrones (see also the vertical line in
Fig.~\ref{fig_mto}).

Finally, we like to remark that the observational $(V-I)$ colors of
 red giant stars are 
tipically 0.1~mag  redder than
predicted by the theory. This small discrepancy is less of a problem
in the $(B-V)$ colour. 
The obvious explanation is
that the transformations from $T_{\rm eff}$ to $(V-I)$ colors still suffer
from some uncertainty.  In recent years, several authors have called
attention on this  problem 
(see Worthey 1994; Gratton et al. 1996; Weiss \& Salaris 1998),
which likely resides in the Kurucz (1992) library of stellar spectra
and associated transformations.

\begin{table}
\caption{Basic calibrations}
\label{tab_deltav}
\begin{tabular}{lllll}
\noalign{\smallskip}\hline\noalign{\smallskip}
 & \multicolumn{2}{c}{Girardi et al.\ 1998} & 
\multicolumn{2}{c}{Bertelli et al.\ 1994} \\
$\log({\rm age/yr})$ & $V_{\rm TO}$ & $\Delta V_{\rm TO}^{\rm RGC}$ &
$V_{\rm TO}$ & $\Delta V_{\rm TO}^{\rm RGC}$  \\
\noalign{\smallskip}\hline\noalign{\smallskip}
9.0 & 2.02 & 0.77 & 2.08 & 1.16 \\
9.2 & 2.68 & 1.70 & 2.68 & 1.79 \\
9.4 & 3.28 & 2.32 & 3.37 & 2.49 \\
9.6 & 3.89 & 2.85 & 3.86 & 2.90 \\
9.8 & 3.83 & 2.72 & 4.20 & 3.16 \\
10.0 & 4.24 & 3.06 & 4.40 & 3.28 \\
10.2 & 4.66 & 3.41 & 4.75 & 3.58 \\
\noalign{\smallskip}\hline\noalign{\smallskip}
\end{tabular}
\end{table}

\subsection{Isochrone fitting method}

Isochrone fitting is to be preferred  to other more
sophisticated methods (e.g. the synthetic CMD technique)  for two
reasons:

\begin{description}

\item $\bullet$ It gives a straightforward  idea of the
best fit age.

\vskip 0.2truecm

\item $\bullet$ It can be used  also when no  membership
for all the  cluster stars is available.

\end{description}

Membership  is a very delicate issue, because luminosity functions and
synthetic CMDs can be successfully used only when the membership is
known.  Indeed  they are based on a quantitative comparison
(number of stars in different evolutionary stages)  between
theory and observations. To assess the  membership of all cluster 
stars is a hard task not always accomplished or feasible  
(Chen et al. 1998, Von Hippel \& Sarajedini 1998).

The cluster metallicities $[Fe/H]$ are translated into theoretical
metal abundances $Z$ by means of the relation:

\begin{equation}
[Fe/H] = log \frac{Z}{0.019} = log Z - 1.721
\end{equation}

\noindent
To derive the age, distance modulus and reddening of a cluster we
proceed as follows.

If RGC stars are present and easy to identify,
the  $\Delta V_{TO}^{RGC}$ method already allows us to select
the appropriate age range  for the cluster under examination.

The main drawback of the 
$\Delta V_{\rm TO}^{\rm RGC}$ method is the identification of the RGC
stars in the CMD of old clusters when the clump gets scarcely
populated. The task is even more difficult if one considers that
open clusters are usually close to the Galactic Plane and therefore are
highly contaminated by field stars.

Once the age range is selected,   with the aid of three isochrones of 
slightly different ages, 
the appropriate values of reddening and distance
modulus are independently derived from the superposition of the
theoretical isochrones onto the observed CMDs.
Distance moduli were selected to provide a good fit of the 
turn-off and subgiant branch magnitude at the same time, and 
reddening was estimated by matching with the isochrone the blue edge
of the main sequence band. 
Finally, only solutions were retained that were able to match the
turn-off and subgiant features but also the position of the RGC stars.

In several cases, it was impossible to
reproduce the CMD both in the $BV$ and $VI$ pass-bands with a single
value  of reddening as governed by the 
relation $E_{(V-I)}= 1.25\,E_{(B-V)}$ (cf. Munari \& Carraro
1996).

\begin{figure*}
\centerline{\psfig{file=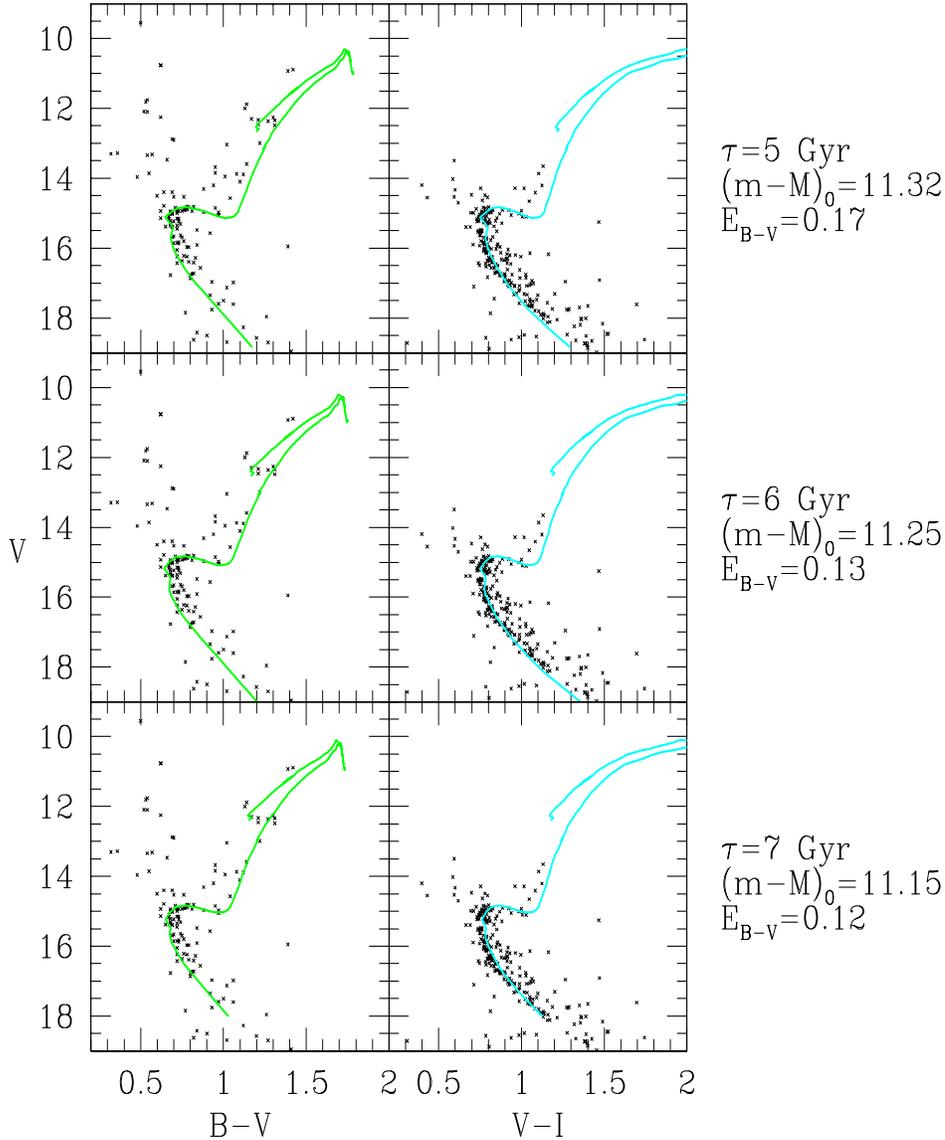,height=18cm,width=14cm}}
\caption{Isochrones superposed to the  CMD of NGC~188.
At the right side we indicate the adopted reddening $E_{(B-V)}$,
distance
modulus $(m-M)_0$, and  age of the isochrone plotted in each 
sub-panel}
\label{fig_188}
\end{figure*}

\begin{figure*}
\centerline{\psfig{file=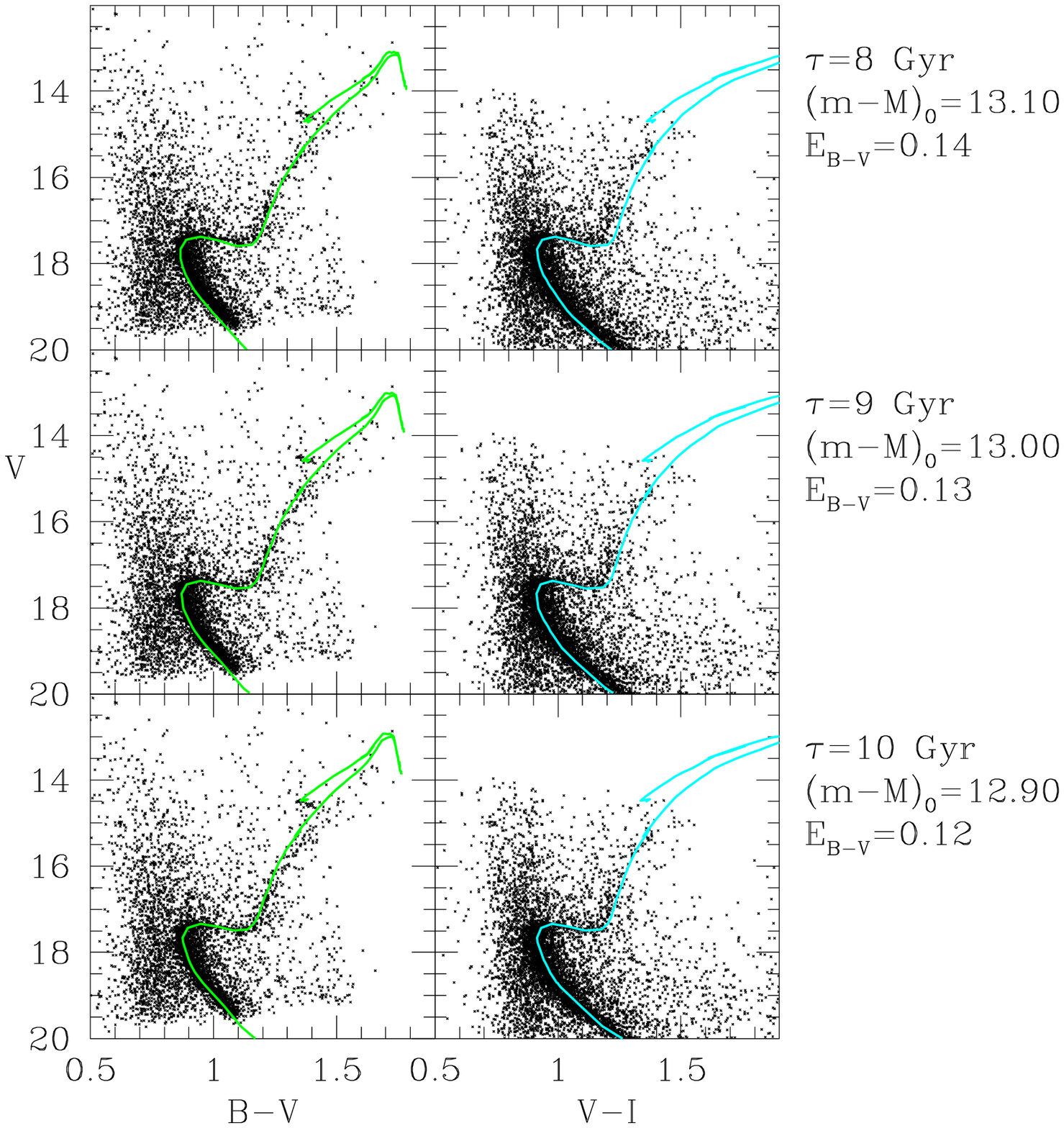,height=18cm,width=14cm}}
\caption{Isochrones superposed to the  CMD of NGC~6791.
At the right side we indicate the adopted reddening $E_{(B-V)}$,
distance
modulus $(m-M)_0$, and  age of the isochrone plotted in each 
sub-panel}
\label{fig_6791}
\end{figure*}

\begin{figure*}
\centerline{\psfig{file=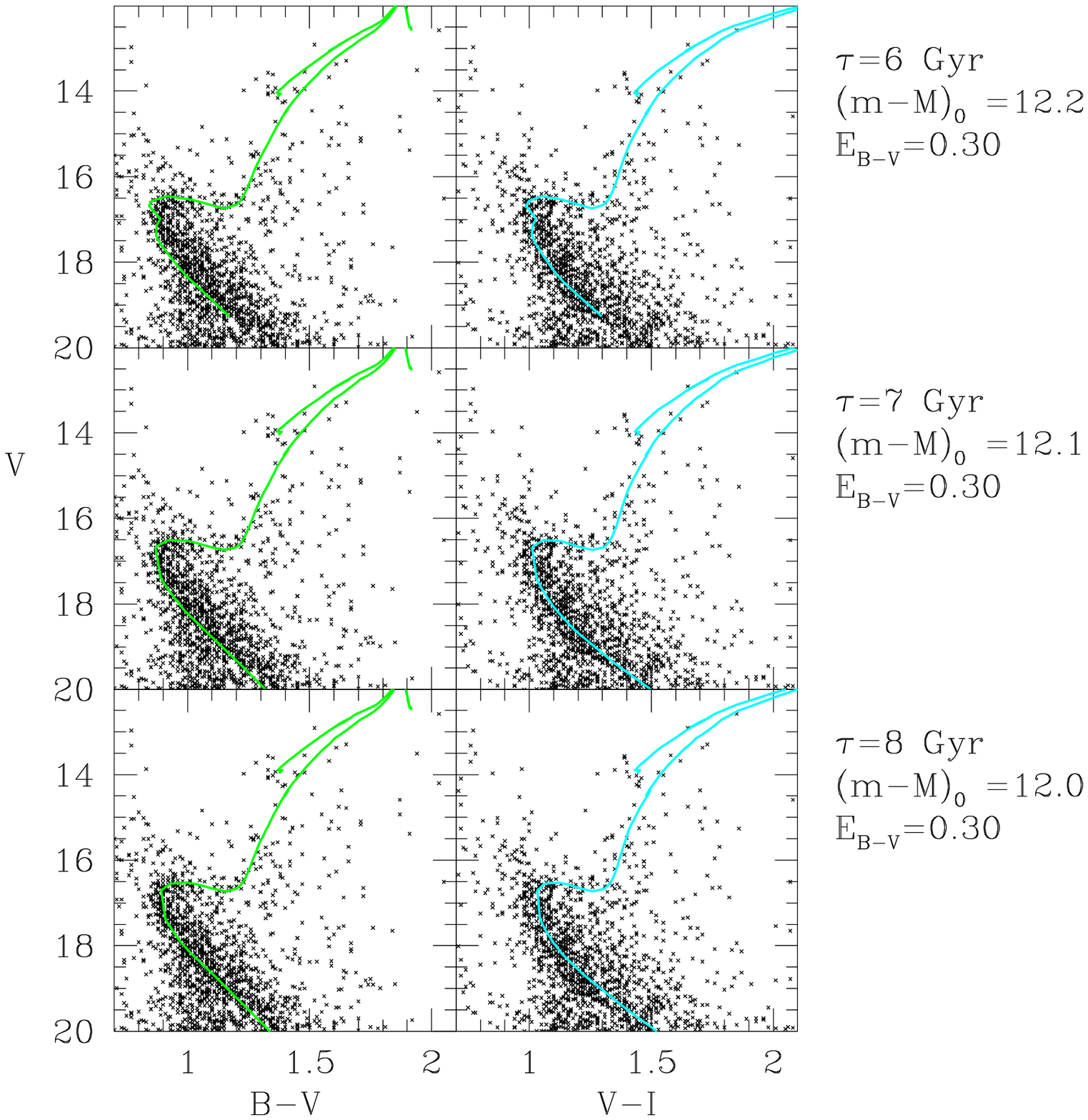,height=18cm,width=14cm}}
\caption{Isochrones superposed to the  CMD of Collinder~261.
At the right side we indicate the adopted reddening $E_{(B-V)}$,
distance
modulus $(m-M)_0$, and  age of the isochrone plotted in each 
sub-panel}
\label{fig_261}
\end{figure*}

\begin{figure*}
\centerline{\psfig{file=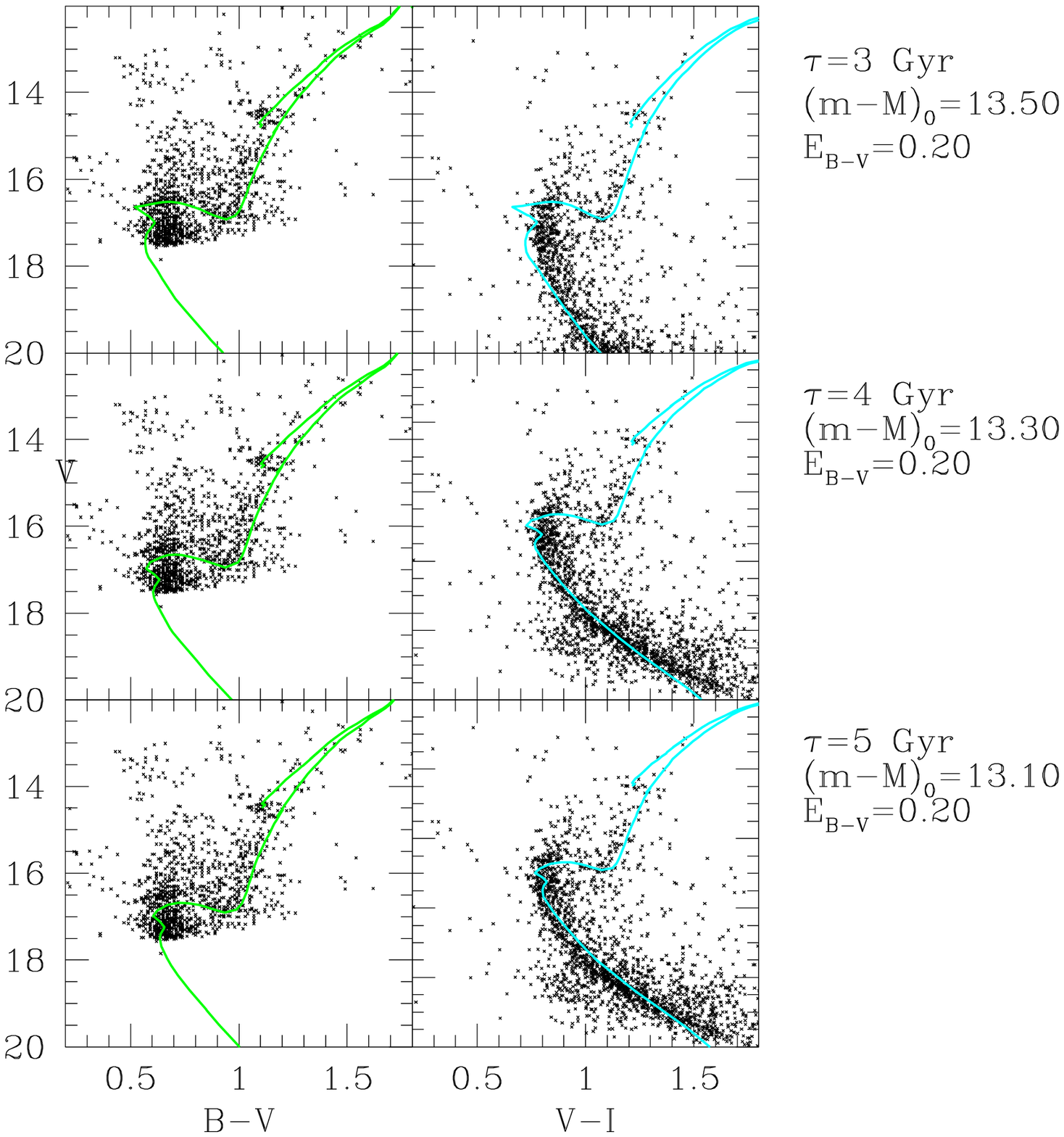,height=18cm,width=14cm}}
\caption{Isochrones superposed to the  CMD of Melotte~66.
At the right side we indicate the adopted reddening $E_{(B-V)}$,
distance
modulus $(m-M)_0$, and  age of the isochrone plotted in each 
sub-panel}
\label{fig_66}
\end{figure*}

\begin{figure*}
\centerline{\psfig{file=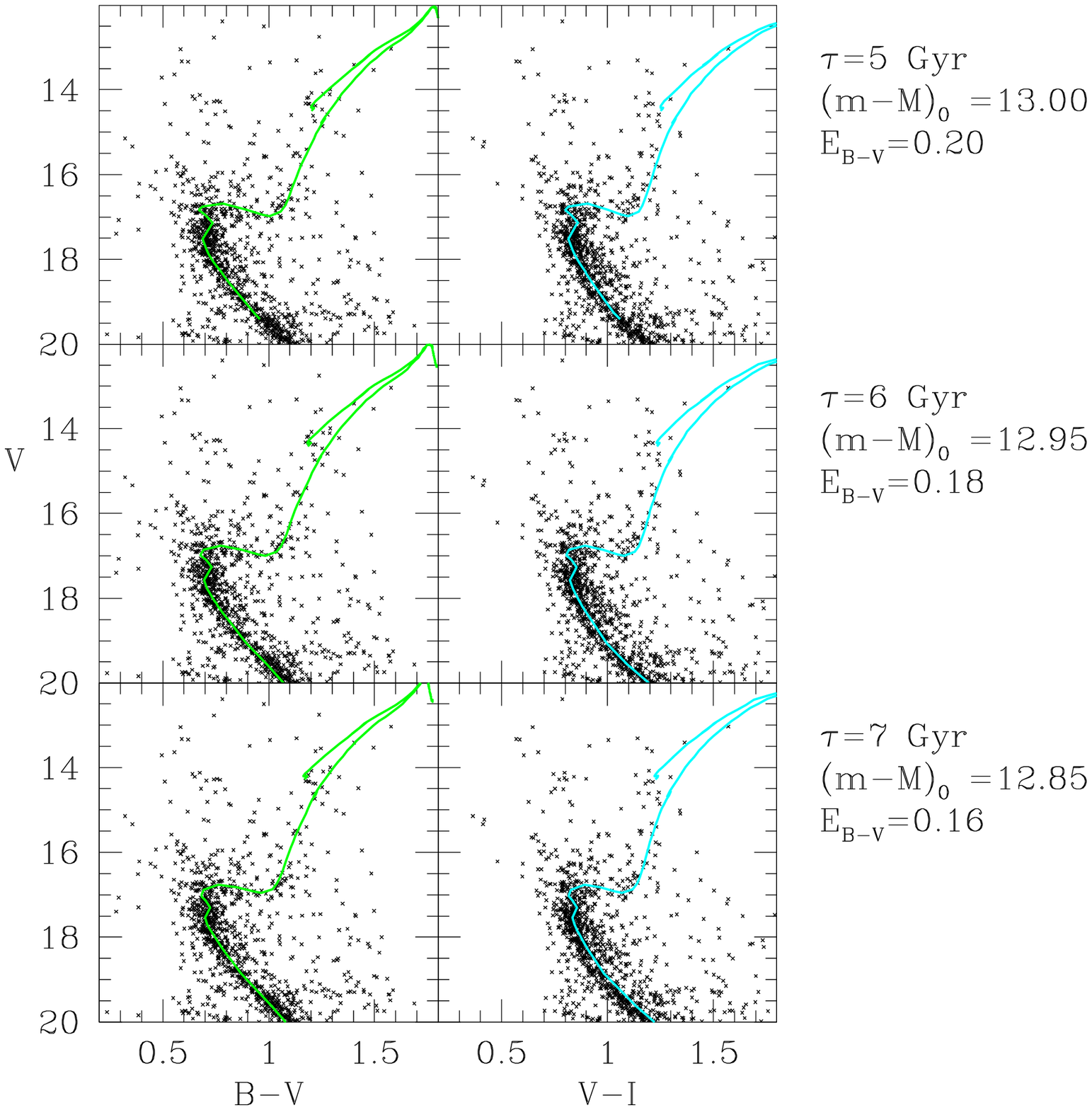,height=18cm,width=14cm}}
\caption{Isochrones superposed to the  CMD of Berkeley~39.
At the right side we indicate the adopted reddening $E_{(B-V)}$,
distance
modulus $(m-M)_0$, and  age of the isochrone plotted in each 
sub-panel}
\label{fig_39}
\end{figure*}

\begin{figure*}
\centerline{\psfig{file=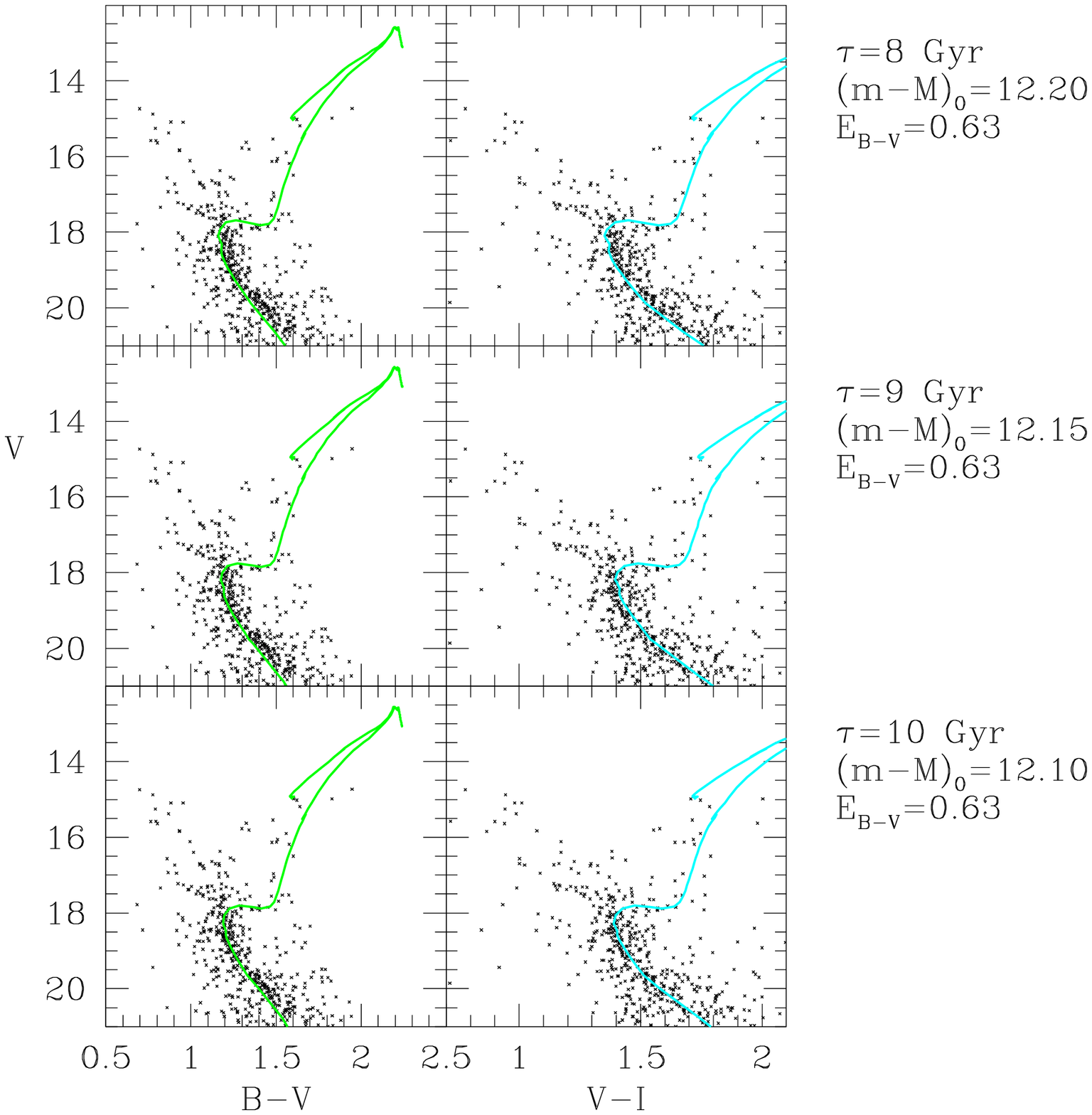,height=18cm,width=14cm}}
\caption{Isochrones superposed to the  CMD of Berkeley~17 by 
Kaluzny (1994).
At the right side we indicate the adopted reddening $E_{(B-V)}$,
distance
modulus $(m-M)_0$, and  age of the isochrone plotted in each 
sub-panel}
\label{fig_17ka}
\end{figure*}

\begin{figure*}
\centerline{\psfig{file=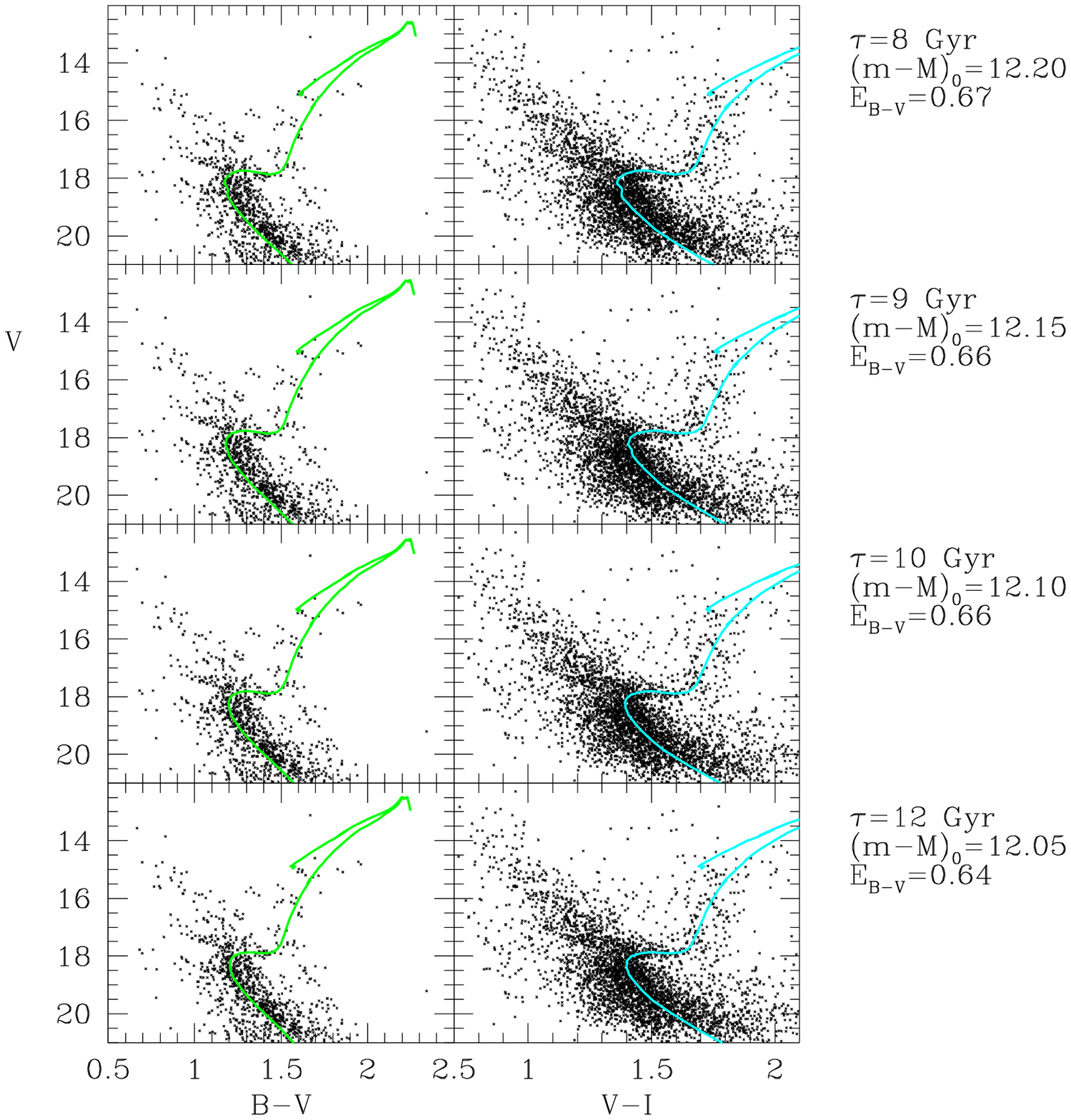,height=18cm,width=14cm}}
\caption{Isochrones superposed to the  CMD of Berkeley~17 by
Phelps (1997), assuming the metallicity value Z = 0.007.
At the right side we indicate the adopted reddening $E_{(B-V)}$,
distance
modulus $(m-M)_0$, and  age of the isochrone plotted in each 
sub-panel}
\label{fig_17phe}
\end{figure*}

\begin{figure*}
\centerline{\psfig{file=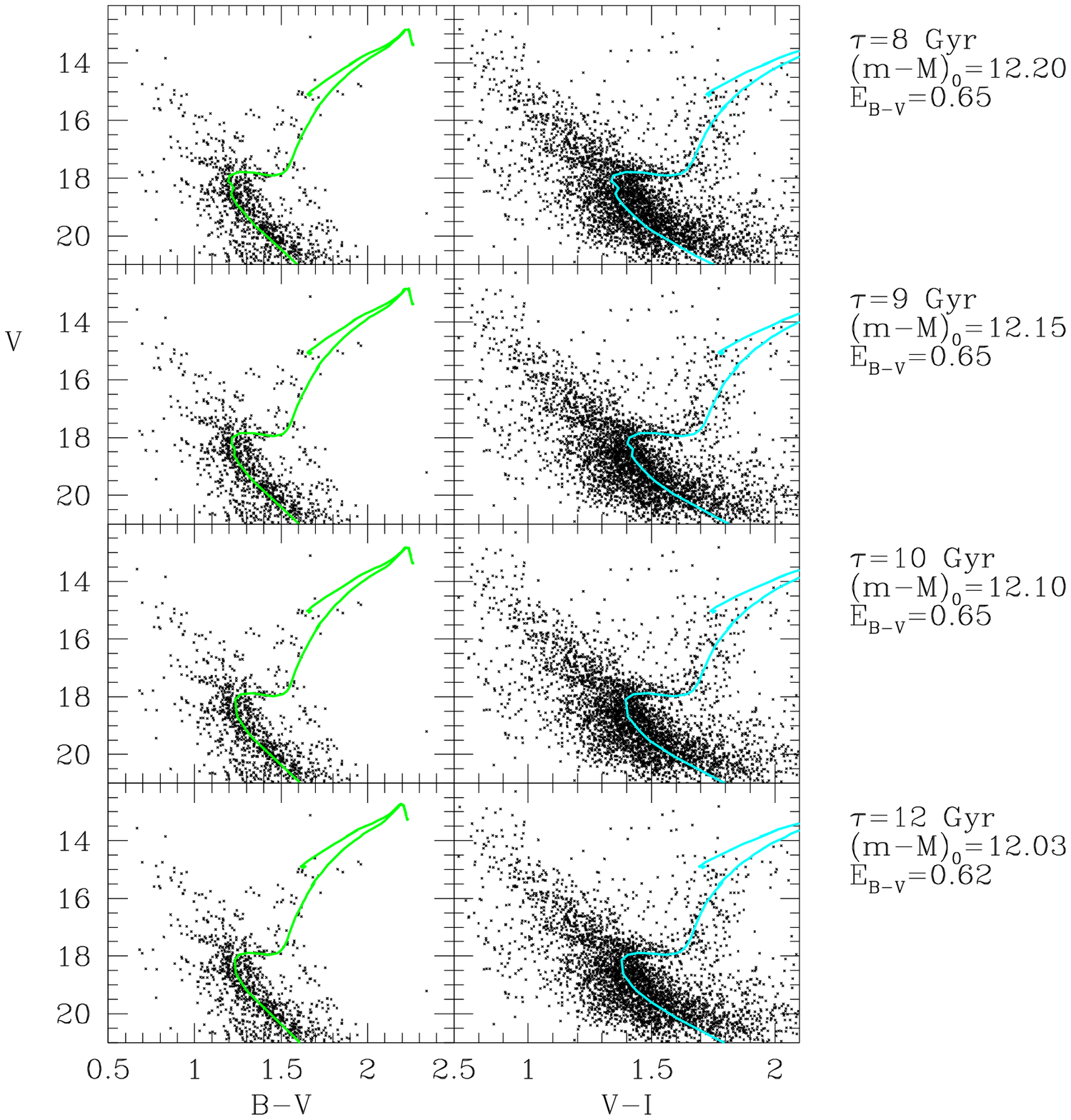,height=18cm,width=14cm}}
\caption{Isochrones superposed to the  CMD of Berkeley~17 by
Phelps (1997), assuming the metallicity value Z = 0.010.
At the right side we indicate the adopted reddening $E_{(B-V)}$,
distance
modulus $(m-M)_0$, and  age of the isochrone plotted in each 
sub-panel}
\label{fig_17phe}
\end{figure*}

\begin{figure*}
\centerline{\psfig{file=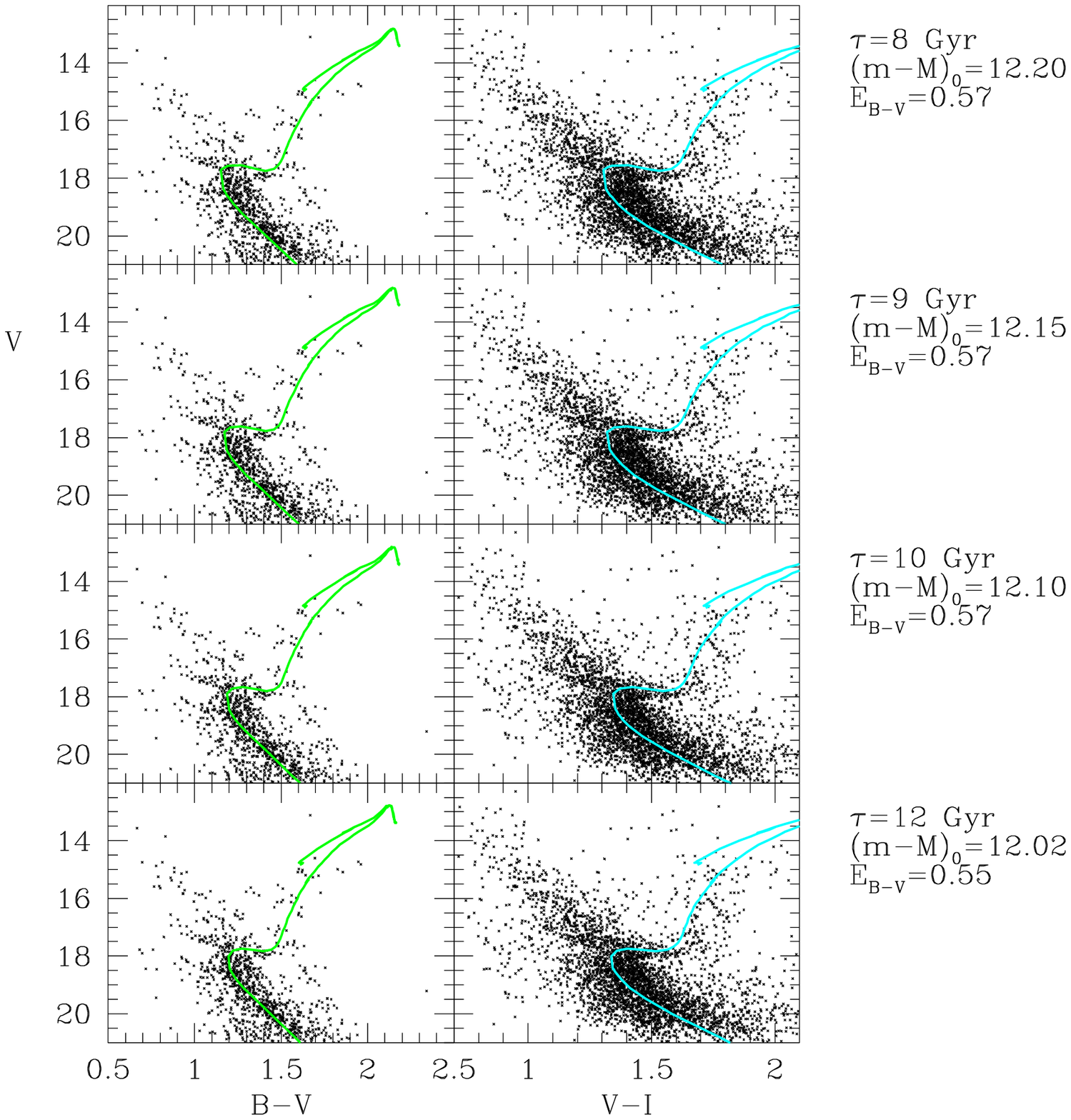,height=18cm,width=14cm}}
\caption{Isochrones superposed to the  CMD of Berkeley~17 by
Phelps (1997), assuming the metallicity value Z = 0.013.
At the right side we indicate the adopted reddening $E_{(B-V)}$,
distance
modulus $(m-M)_0$, and  age of the isochrone plotted in each 
sub-panel}
\label{fig_17phe}
\end{figure*}

\section {Cluster by cluster analysis}

For each cluster listed in Table~1, we derive the age comparing  the
observational CMD with 
isochrones of given metallicity. The analysis is simultaneously made
both  in  the $V$ vs.\ $B-V$ and $V$ vs.\ $V-I$
diagrams.

The results  are  summarized in Table~3, which 
for each cluster lists the metallicity $Z$, the color excess $E_{(B-V)}$ and
$E_{(V-I)}$, the distance moduls $(m-M)_0$, the age $\tau$ (in Gyr),
together with their uncertainties.

\begin{table*}
\tabcolsep 0.35truecm 
\caption{Estimated reddening, distance modulus, and age for  our sample of  open clusters}
\begin{tabular}{cccccc} \hline
\multicolumn{1}{c}{Cluster} &
\multicolumn{1}{c}{$Z$ }&
\multicolumn{1}{c}{$E_{(B-V)}$} &
\multicolumn{1}{c}{$E_{(V-I)}$} &
\multicolumn{1}{c}{$(m-M)_o$} &
\multicolumn{1}{c}{$\tau$ (Gyr)}  \\
\hline
NGC~188       &0.017& $0.12-0.13$ & $0.15-0.16$& $11.20\pm 0.05$ & $6-7$ \\
NGC~6791      &0.030& $0.13-0.14$ & $0.15-0.16$& $13.05\pm 0.05$ & $8-9$ \\
Collinder~261 &0.014&  0.30           & 0.36           & $12.15\pm0.05$  & $6-7$ \\
Melotte~66    &0.006&  0.20           & 0.25           & $13.20\pm 0.10$ & $4-5$ \\
Berkeley~39   &0.009& $0.18-0.20$ & $0.22-0.25$& $12.97\pm 0.02$ & $5-6$ \\   
Berkeley~17   &0.007& $0.64-0.67$ & $0.73-0.78$& $12.13\pm 0.07$ & $8-10$\\
Berkeley~17   &0.010& $0.63-0.65$ & $0.78-0.81$& $12.13\pm 0.07$ & $8-10$\\   
Berkeley~17   &0.013& $0.55-0.57$ & $0.64-0.68$&  $12.13\pm 0.07$ & $8-10$\\
\hline
\hline
\end{tabular}
\end{table*}

\subsection{NGC~188}
The source of photometric data for NGC~188 is  the $BV$ photographic survey of
McClure \& Twarog (1977)  and  the $VI$ CCD survey of Von Hippel \&
Sarajedini (1998), 
whereas the metal abundance $[Fe/H]$ is  from Friel et al. (1995).\\
The comparison of the CMD with  isochrones for $Z=0.017$ is shown in 
Fig.\ref{fig_188}.
The best fit age is  derived  from the $B$ vs. $(B-V)$ CMD, in which
the clump of He-burning stars is located at 
 $V \approx 12.2$ and $(B-V \approx 1.3$). This group of stars is missing
in the  $V$ vs. $(V-I)$ CMD because of the poor statitistics.  
We get an age of $6-7$ Gyr, a distance modulus $(m-M)_0=11.20 \pm
0.05$, and
a color excess $E_{(B-V)}=0.12 - 0.13$. The solution with age of 5
Gyr, $(m-M)_0=11.32$, $E_{(B-V)}=0.17$ can be discarded because of the
poor fit of the red clump stars.  Within the observational uncertainties, 
age,  distance modulus, and reddening we have obtained are consistent 
with published values (see Carraro \& Chiosi 1994; Carraro et
al. 1994;   Von Hippel \& Sarajedini 1998).

\subsection{NGC~6791}
NGC~6791 has received much attention over the  past years because of 
its unique combination of age and metallicity.
In addition to  being one of the most populous open clusters, 
it is the most metal-rich and among the  oldest clusters at the same time. 
It contains also seven blue horizontal
branch stars.\\
We adopt the photometric data by Kaluzny \& Rucinski 
(1995), and the spectroscopic metallicity by Friel \& Janes (1993).
Our best fit is shown in Fig.~\ref{fig_6791}. 
We get an age of $8- 9$ Gyr, a distance modulus $(m-M)_0=13.05 \pm
0.05$, and 
a color excess $E_{(B-V)}=0.13- 0.14$. The solution with age of 10
Gyr, $(m-M)_0=12.90$, and  $E_{(B-V)}=0.12$ is  rejected  because of the
poor fit of the red clump stars.  
Also in this case, age, distance modulus and reddening agree with
previous estimates, e.g. the age of 9 Gyr quoted by 
Carraro et al. (1994a) and Garnavich et al. (1994). It is
significantly older than the 8 Gyr quoted by Carraro \& Chiosi (1994).
 The age is
marginally
older than that found by Carraro et al. (1994a), which is 
due to the systematic difference of 0.15 mag passing from Bertelli et
al. (1994) to Girardi et al. (1998) isochrones.

\subsection{Collinder~261}
This is a populous cluster studied by Mazur et al. (1995) and Gozzoli et
al. (1996), who present photometric data of similar  quality. We adopt
here the  CMDs of   Mazur et al. (1995).
The metallicity $[Fe/H]$
is derived from the spectroscopic study of Friel et al. (1995).
The superposition of theoretical isochrones onto the observational CMDs
is shown in Fig.~\ref{fig_261}. 
The age cannot be estimated better than  $6- 8$ Gyr,
even
though there is a marginal preference for the narrower range $6- 7$
Gyr, which is reported in Table~3. The distance modulus is 
$(m-M)_0=12.15 \pm 0.05$,
and the color excess is $E_{(B-V)}=0.30$. 
In this case, our age determination is
fully consistent
with that by Mazur et al. (1995), but only marginally  with
the one by Gozzoli et al. (1996). Reddening and distance modulus 
are however in
agreement with both previous studies. The cause for the  age 
disagreement can be attributed  to differences in the adopted
stellar models and technique to generate isochrones and synthetic CMDs.

\subsection{Melotte~66}
$B$ and $V$ photometry for Melotte~66 is available only from the 
photographic study of Anthony-Twarog et al. (1979) which is  deep
enough to reach the turn-off. In contrast the $V$ and $I$ photometric
data by  Kassis et al. (1997) go much deeper.
The cluster is the most metal poor of the sample (Friel \& Janes
1993).
The comparison with the isochrones is shown in Fig.~\ref{fig_66}. 
We get an age of $4- 5$ Gyr, a distance modulus $(m-M)_0=13.20 \pm
0.10$, and
a color excess $E_{(B-V)}=0.20$. The solution with age of 3
Gyr, $(m-M)_0=13.5$, and $E_{(B-V)}=0.20$ is  rejected  because of the
poor fit of the red clump stars.  
Melotte~66 turns out
to be the youngest cluster of the sample. 
The distance modulus and reddening found for  Melotte~66 agree with  the 
detailed analysis of Kassis et al. (1997).

\subsection{Berkeley~39}

The  
photometric data for this cluster  is  from
Kassis et al. (1997), whereas the metal abundance is  from Friel \&
Janes (1993).
The CMD of Berkeley~39 and the comparison with theoretical isochrones 
is shown in Fig.~\ref{fig_39}.
The best fit case is for age  $5- 6$ Gyr,   distance modulus
$(m-M)_0=12.975 \pm
0.025$, and
 color excess $E_{(B-V)}=0.18- 0.20$. The solution with age of 7
Gyr, $(m-M)_0=12.85$, and $E_(B-V)=0.16$ is  rejected  because of the
poor fit of the turn-off and clump of red  stars.  
The new parameters are significantly different from what found by Carraro \&
Chiosi (1994), who gave $E_{(B-V)}=0.10$, $(m-M)_0=13.20$, and age of 6.5
Gyr, whereas within the uncertainties
reddening and distance modulus are  consistent 
  with the determinations by Kassis at al. (1997).

\subsection{Berkeley~17}
The analysis carried in so far  clearly shows  that  ages
based the new set of isochrones (Girardi et al. 1998) are  
comfortably  consistent with previous estimates. This is an important
remark to be made in view of the results we would obtain for the
most controversial cluster in the sample, namely Berkeley~17.

This cluster has been studied several times in the optical (Kaluzny
1994;  Phelps et al.
1994, 1995; Phelps 1997) and more recently also in the near IR 
(Carraro et al. 1998b).

Kaluzny (1994) compared Berkeley~17 with NGC~6791 and, assuming that the
former is more metal-poor than the latter, concluded that the two clusters  
are likely coeval (about 9 Gyr old). Moreover, he suggested a  
distance modulus $(m-M)_0 \geq 13.26$ and a reddening $E_{(B-V)} \geq
0.56$ or $E_{(V-I)} \geq 0.70$.

In contrast, Phelps (1997) using photometric data much similar
to that by Kaluzny (1994) reached considerably different conclusions,
Specifically,  he found a metallicity in the range $-0.30 \leq [Fe/H]
\leq 0.00$,  an age of
$12^{+1}_{-2}$ Gyr, a distance modulus $(m-M)_0 \approx 12.15$ and a 
reddening in the range $0.52~<~E_{(B-V)}~<~0.68$ and
$0.61~<~E_{(V-I)}~<0.71$.

In this study we adopt the metallicity  by   Friel et al. (1995), 
i.e. $[Fe/H]= -0.29\pm 0.13$,
to which corresponds the a metallicity in the in the range $Z=0.007-0.013$.  
We explored this metallicity range performing fits for Z=0.007, 0.010
and 0.013 (see Figs. 9,10 and 11, respectively).
Good fits
of the CMDs from both sources of data, shown in 
Figs.~\ref{fig_17ka} (Kaluzny 1994) and \ref{fig_17phe} (Phelps 1997),
are possible for the following combinations of parameters: 
$E_{(B-V)}=0.55- 0.67$
$E_{(V-I)}=0.64-0.78$, $(m-M)_0=12.13 \pm0.07$, and age $9\pm 1$
Gyr. 
The effect of changing the metallicity does not affect the age determination,
but only the value of the reddening.

An age of 12 Gyr is clearly ruled out for any metallicity, since fitting the turn-off
the theoretical clump turns out to be too bright.

Our analysis stands on the identification of the RGC stars
as  the handful of objects located at $V \simeq 15.2$ and $B-V \simeq 1.65$ or
$(V-I) \simeq 1.72$ (Phelps 1997) and the simultaneous fit of the
turn-off,  subgiant and RGC stars.  
The  results we get  for the cluster parameters   are also   consistent 
with companion analysis of the CMD in the near IR (J and K pass-bands)
by  Carraro et al. (1998b).

{\it
Why the new age is younger than what found by Phelps (1997) using the
same data, reddening,  and distance modulus? }

Phelp's (1997) result is even more surprising if one considers that 
passing from Bertelli's et al. (1994) to Girardi's et al. (1998)
isochrones a systematic {\it increase}  in the age of about 19\% for old 
clusters like Berkeley~17 is expected.

Owing to the many implications deriving from an age for Berkeley~17
as old as that of the bulk of globular clusters, a close scrutiny
of the Phelps (1997) study is required. 

His analysis proceeds in two steps. Firstly the CMDs are compared with
the VandenBerg (1985) isochrones (they  do not extend beyond the 
subgiant branch) and no assumptions are made for the metallicity,
reddening, and distance. Out of this first comparison the conclusion is
drawn that $[Fe/H]=-0.23$ and age of 12.5 Gyr best match the $BV$ and
$VI$ data. The analysis is then repeated using the Bertelli et
al. (1994) isochrones for ages of 10.0,  12.0 and 13.2 Gyr and
metallicities $Z=0.05$, $Z=0.02$ and $Z=0.008$. In this second
approach, the ability of the various isochrones in matching the
location of RGC stars is also taken into account. Looking at the series
of Figs.7, 8, and 9 in Phelps (1997), we would surely exclude
all cases with age 13.2 Gyr and also all cases with Z=0.05 and 
marginally Z=0.020 as well,  because the RGC stars are not matched.
In contrast, the cases with $Z=0.008$ and ages of 10.0 and 12.0 Gyr
(to a less extent)
provide a satisfactory fit of all the constraints. This is actually 
what find in our analysis: $Z\simeq 0.010$ and age $9\pm 1$ Gyr. 
Somehow rejecting his own results and perhaps influenced by the
fit with the VandenBerg (1985) isochrones, Phelps (1997) preferred to
conclude that the age of Berkeley~17 is $12^{+1}_{-2}$ Gyr.
In contrast, the age that we would  derive from Phelps' (1997)
study is   $11^{+1}_{-1}$ Gyr. The younger
age comes from the fainter RGC stars in the Girardi et al. (1994)
isochrones.

\section {Conclusions}
In this paper we have derived the ages of a sample of very old open
clusters.
The results are summarized in
Table~3. The  ages of the clusters under examination range from 
 $4-5$ Gyr  (Melotte~66) to $8-9$ Gyr
(NGC~6791) and $9\pm 1$ Gyr (Berkeley~17). 

Particularly intriguing  is the age we have
found for Berkeley~17 ($9\pm 1$ Gyr). While the new age fairly agrees with
the
estimate of 9 Gyr suggested by Kaluzny (1994), it is probably younger 
than the $12^{+1}_{-2}$ Gyr  claimed by Phelps (1997) and also the 
revised value of  $11^{+1}_{-1}$ Gyr we have suggested. The reason for the
difference is found in the use of different sets of 
isochrones that differ in some important details.

According to the present analysis 
the age of Berkeley~17 is no
longer embarrassingly close
to that of bulk globular clusters: 13 Gyr with a tail down to 11 Gyr
(Gratton et al. 1997), but most likely falls back into the classical  range for
this type of clusters. However
an old age such as  that found by Phelps (1997) 
-- we prefer to consider here  the value of
$11^{+1}_{-1}$ Gyr -- cannot be firmly excluded because of the many
uncertainties  still affecting the stellar models in the mass range
1.0 to 1.5 $M_{\odot}$.

Perhaps, the most important conclusion to be learned from the present
study is that  we are
facing the embarrassing situation
in which our poor knowledge of important details of the physical
 structure of low mass stars in the range 1.0 to  1.5 $M_{\odot}$, 
for instance
central mixing and associated overshooting,
does not allow us to derive ages of old open clusters with the
required precision.  Therefore, to answer the basic
question posed by the maximum age of old open clusters with respect to
that of globular clusters, one has to wait until this point is
 clarified. A thorough investigation of the behaviour of convective
 cores
in the mass range 1.0 -1.5 $M_{\odot}$ is a prerequisite to age
studies  of any kind.  

Despite this serious drawback of extant theory of stellar structure
and evolution let us speculate further on the provisional assumption
that
the ages we have found are not too grossly in error.

{\it If Berkeley~17 and NGC~6791 trace the upper limit to the age of 
open clusters, how this apply to the age of the  
Galactic Disk?  Are there other  not yet discovered  open clusters of 
older age? If  open clusters set the a sort of limit to the age of the
Galactic Disk, is this value 
consistent with other independent age estimates of this component of
the Milky Way?}

A variety of different methods can be used to derive the age of the 
Galactic Disk.
In summary, the situation is as follows: 

\begin{description}
\item $\bullet$  The luminosity function  of White Dwarfs suggests an 
age between 6 and 10 Gyr (Bergeron et al. 1997).
 
\item $\bullet$  The oldest evolved G and F stars in the Edvardsson et
al. (1993) sample  have ages around 10 Gyr according to
the analysis made by Ng \& Bertelli (1997). 

\item $\bullet$ Radio-active dating (from isotope ratios) gives
estimates around 9 Gyr (Morell et al. 1992); 

\item $\bullet$ The faintest sequence of subgiant stars 
suggests an age of
about $11 \pm 1$ Gyr (Jimenez et al. 1998).

\end{description}

Adding together all these different estimates, we can argue that the Galactic Disk 
has an age of about 9-10 Gyr, 2-3 Gyrs younger than the bulk of globular clusters
(13 Gyrs, Gratton et al. 1997).
The hint arises that the Milky Way underwent a minimum activity or
hiatus in its Star Formation History in the provisional age range  10 to 11  Gyrs ago.

\section*{Acknowledgements}
G.C. acknowledges useful discussions with drs Y.K. Ng and A. Bressan, and the referee,
dr Randy Phelps for his useful suggestions which helped to improve on the paper.
C.C. wishes  to acknowledges the hospitality and
stimulating environment provided by MPA in Garching where this paper
has been completed  during  leave of absence from the Astronomy Department of
the Padua University.
This study has been financed by the Italian Ministry of
University, Scientific Research and Technology (MURST) and the Italian
Space Agency (ASI).
The work of L\'eo Girardi is funded by the Alexander von Humboldt-Stiftung.


\begin{thebibliography}{}

\bibitem{} 
Alexander D.R, Ferguson J.W., 1994, ApJ 437, 879

\bibitem{} 
Alongi M., Bertelli G., Bressan A., Chiosi C., Fagotto F., Greggio L.,
Nasi E., 1993, A\&AS 97, 851


\bibitem{}
Anthony-Twarog B.J., Twarog B.A., McClure R.D., 1979, ApJ 233, 188

\bibitem{}
Bergeron J., Ruiz M.T., Legget S.K., ApJS 108, 339

\bibitem{}
Bertelli G., Bressan A., Chiosi C., Fagotto F., Nasi E., 1994, A\&AS 106, 275

\bibitem{}
Bertelli G., Bressan A., Chiosi C., 1992, ApJ 392, 522

\bibitem{}
Bressan A., Chiosi C., Fagotto F., 1994, ApJS 94, 63

\bibitem{}
Cannon R.D., 1970, MNRAS 150, 111

\bibitem{}
Carraro G., Chiosi C., 1994, A\&A 287, 761


\bibitem{}
Carraro G., Chiosi C., Bertelli G., Bressan A., 1994, A\&AS 103, 375

\bibitem{}
Carraro G., Girardi L., Bressan A., Chiosi C., 1996, A\&A 305, 849

\bibitem{}
Carraro G., Ng Y.K., Portinari L., 1998a, MNRAS 296, 1045

\bibitem{}
Carraro G., Vallenari A., Girardi L., Richichi A., 1998b, A\&A 343, 825

\bibitem{}
Chen B., Carraro G., Torra J., Jordi C., 1998, A\&A 331, 916

\bibitem{}
Chiosi C., Bertelli G., Bressan A., 1992, ARA\&A, 30, 235

\bibitem{}
Edvardsson B., Andersen J., Gustafson B., Lambert D.L., Nissen P., 
Tomkin J., 1993, A\&A 275, 101

\bibitem{} 
Friel E.D., Janes K.A., 1993, A\&A, 267, 75

\bibitem{} 
Friel E.D., 1995, ARA\&A, 33, 381

\bibitem{}
Friel E.D., Janes K.A., Hong L., Lotz J., Tavarez M., 1995, in {\it
The Formation of the Milky Way}, eds. E.J. Alfaro and A.J. Delgado, 
p. 189,  Cambridge University Press

\bibitem{}
Garnavich P.M., VandenBerg D.A., Zurek D.R., Hesser J.E., 1994, AJ 107, 1097

\bibitem{} 
Girardi L., Bressan A., Chiosi C., 1996, in Stellar Evolution: 
        What Should Be Done, 32nd Li\`ege Int.\ Astrophys.\ Coll.,
        eds.\ A.\ Noels et al., p.\ 39.

\bibitem{}
Girardi L., Bressan A., Bertelli G., Chiosi C., 1998, in preparation

\bibitem{}
Gozzoli E., Tosi M., Marconi G., Bragaglia A., 1996, MNRAS 283, 66

\bibitem{}
Gratton R.G., Carretta E., Castelli F., 1996, A\&A 314, 191

\bibitem{}
Gratton R.G., Fusi Pecci F., Carretta G., Clementini G., Corsi C.E., 
Lattanzi M., 1997, ApJ 491, 749

\bibitem{}
von Hippel T., Sarajedini A., 1998, AJ 116, 1789

\bibitem{}
Iglesias C.A., Rogers F.J., 1996, ApJ 464, 943

\bibitem{}
Janes K.A., 1989, in {\it Calibration of Stellar Ages}, ed A.G. Davis
Philip, p. 59,   L. Davis press, Schenectady, N.Y.

\bibitem{}
Janes K.A., Phelps R.L.., 1994, AJ 108, 1773

\bibitem{}
Jimenez R., Flynn C., Kotoneva E., 1998, MNRAS 299, 515

\bibitem{}
Kaluzny J., 1994, AcA 44, 247 

\bibitem{}
Kaluzny J., Rucinski S.M., 1995, A\&AS 114, 1

\bibitem{}
Kassis M., Janes K.A., Friel E.D., Phelps R.L., 1997, AJ 113, 1723

\bibitem{} 
Kurucz R.L., 1992, in The Stellar Populations of Galaxies, 
        eds. B.\ Barbuy and A.\ Renzini, p. 225,  Dordrecht: Kluwer   

\bibitem{}
McClure R.D., Twarog B.A., 1977, ApJ 219, 111

\bibitem{}
Morell O., K\"allander D., Butcher H.R., 1992, A\&A 259, 543

\bibitem{}
Munari U., Carraro G., 1996, A\&A 314, 108

\bibitem{}
Ng Y.K., Bertelli G., 1997, A\&A 329, 943

\bibitem{}
Olive K., Steigman G., 1995, ApJS 97, 49

\bibitem{}
Perryman M.A.C., Brown A.G.A., Lebreton Y., Gomez A., Turon C., 
De Strobel G.C., Mermilliod
J.C., Robichon N., Kovalevsky J., Crifo F., 1998, A\&A 331, 81

\bibitem{}
Phelps R.L., 1994, Janes K.A., Montgomery K.A., 1994, AJ 107, 1079

\bibitem{}
Phelps R.L., 1994, Janes K.A., Friel E.D., Montgomery K.A., 1995, in
{\it The Formation of the Milky Way}, eds. E.J. Alfaro and
A.J. Delgado, p. 189,  Cambridge University Press

\bibitem{}
Phelps R.L., 1997, ApJ 483, 826

\bibitem{} Reimers D., 1975, Mem.\ Soc.\ R.\ Sci.\ Li\`ege, 
        ser.\ 6, vol.\ 8, p.\ 369 

\bibitem{}
Sandage A.R., 1989, in {\it Calibration of Stellar Ages}, ed.  A.G. 
Davis Philip, p. 43,   L. Davis Press, Schenectady, N.Y.

\bibitem{}
Tavarez M., Friel E.D., 1995, AJ 110, 223

\bibitem{}
Torres-Peimbert S., Peimbert M., Fierro J., 1989, ApJ 345, 186

\bibitem{}
Weiss A., Salaris M., 1998, A\&A, in press

\bibitem{}
Worthey G., 1994, ApJS 95, 107


\end{thebibliography}
\end{document}